\documentclass[twocolumn,runningheads]{svjour2}
\smartqed  % flush right qed marks, e.g. at end of proof
\usepackage{graphicx}
\newcommand{\ltsim}{\raisebox{-.5ex}{$\;\stackrel{<}{\sim}\;$}}

\journalname{Astrophysics and Space Science}
\begin{document}

\title{Our distorted view of magnetars: application of the Resonant Cyclotron Scattering model\thanks{NR is supported by an NWO Post-Doctoral Fellowship. SZ thanks the Particle Physics and Astronomy Research Coucil, PPARC, for support through  an Advanced Fellowship.}}

%\titlerunning{Short form of title}        % if too long for running head

\author{Nanda Rea         \and
        Silvia Zane  \and
        Maxim Lyutikov \and
        Roberto Turolla
}

\authorrunning{Rea, Zane, Lyutikov \& Turolla} % if too long for running head

\institute{N. Rea \at
              SRON Netherlands Institute for Space Research, \\
              Sorbonnelaan, 2, 3584CA, Utrecht, The Netherlands \\
              \email{N.Rea@sron.nl}                 
\and
           S. Zane \at
              Mullard Space Science Laboratory, 
\\University College of London, \\
Holbury St. Mary, Dorking Surrey, RH5 
6NT, UK 
\and
           M. Lyutikov \at
              University of British Columbia, \\
          6224 Agricultural Road, Vancouver, BC, V6T 1Z1, Canada 
\and 
           R. Turolla \at
           University of Padua, Physics Department, \\
       via Marzolo 8, 35131, Padova, Italy }

\date{Accepted: 2006 August 30}
% The correct dates will be entered by the editor

\maketitle

\begin{abstract}
The X-ray spectra of the magnetar candidates are customarily fitted
with an empirical, two component model: an absorbed blackbody and a
power-law. However, 
 the physical interpretation of these two
spectral
 components is rarely discussed. It has been recently
proposed that the presence of a hot plasma in the magnetosphere of
highly magnetized neutron stars
 might distort, through efficient
resonant cyclotron scattering, 
 the thermal emission from the neutron
star surface, 
 resulting in production of non-thermal spectra. 
 Here
we discuss the Resonant Cyclotron Scattering (RCS) model, and present
its XSPEC implementation, as well as preliminary results of its
application to Anomalous X-ray Pulsars and Soft Gamma-ray Repeaters.

\keywords{Neutron Stars \and Pulsars \and Magnetars \and X-ray \and Resonant Cyclotron Scattering}
\PACS{97.60.Jd \and 97.60.Gb}

\end{abstract}

\section{Introduction}
\label{intro}
Anomalous X-ray Pulsars (AXPs) and Soft Gamma-ray Repeaters (SGRs) are
a small class of slowly rotating (5-12 s) neutron stars with emission
properties much at variance with those of ordinary X-ray pulsars, both
the young radio pulsars and the X-ray binary pulsars. They are called
``anomalous'' because their high X-ray luminosity ($10^{34} - 10^{36}$
erg/s) cannot be easily explained in terms of the conventional
processes which apply to other classes of pulsars, i.e. accretion from
a binary companion or injection of rotational energy in the pulsar
wind/magnetosphere. On the other hand, measurements of spin periods
and period derivatives, when the latter are interpreted as due to
electromagnetic dipolar losses, suggest that these objects may host
``magnetars'', i.e. neutron stars endowed with an ultra-strong
magnetic field ($10^{14} - 10^{15}$\,G, see Duncan \& Thompson 1992;
Thompson \& Duncan 1993, 1996). The magnetar scenario appears so far
very promising in explaining both the main energy source of these
objects (the decay of the super-strong field) and the emission of the
short, energetic bursts. Moreover, the magnetar scenario can account
for the properties of Giant Flares, extremely energetic transient
events ($L\approx 10^{44}-10^{47}$ erg/s) detected from SGRs.
However, alternative scenarios to explain the enigmatic properties of
these sources have been invoked. Among these, models involving
accretion from a fossil disk, left by the supernova event which gave
birth to the neutron star, are still largely plausible (van Paradijs
et al. 1995; Chatterjee, Hernquist \& Narayan 2000). Very recently
such a (maybe passive) disk has been indeed observed in the IR
around AXP 4U 0142 + 61 (Wang, Chakrabarty \& Kaplan 2006).  Magnetars
candidates are strong X-ray sources, and their X-ray spectra in the
0.5-10~keV band are usually fit with a two component model consisting
of (besides interstellar absorption) a thermal blackbody with a typical
temperature $kT\sim$0.3--0.4 keV and a power-law with a relatively
steep photon index, $\Gamma\sim$2--4 (see Kaspi 2006 in this volume,
and Woods \& Thompson 2006 for recent reviews). In some cases, SGR
spectra have been fit with a single power-law component, but recent
longer observations showed that, also for these sources, a blackbody
component is often required (Mereghetti et al. 2005).

Despite the fact that the blackbody plus power-law spectral model has
been largely applied to magnetar spectra for many years, a reliable
physical interpretation of these two components is still
missing. Quite recently, an attempt to interpret the observed spectrum
and long term variations of 1E 1048.1--5937 in terms of a (non
resonant) Comptonization model has been presented by Tiengo et
al. (2005).  Furthermore, the recent discovery of magnetar
counterparts in the radio to the $\gamma$-ray bands (Camilo et
al. 2006; Hulleman et al. 2000; Kuiper et al. 2004) enforced the idea
that their multiwavelength spectral energy distribution is by far more
complicated than a simple blackbody plus power-law distribution (see
also Den Hartog et al. contribution in this volume).

%%%%%%%%%%  Figure 1 %%%%%%%%%%%%%%%%%%%%%%%%%%%%%%%%%%%%%
\begin{figure}
\centering
\includegraphics[height=0.35\textwidth,width=0.45\textwidth]{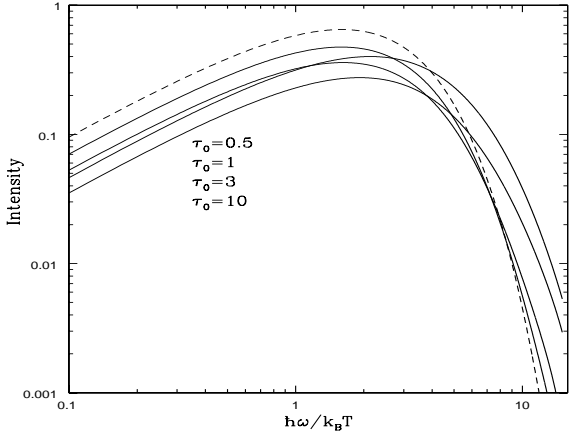}
\caption{Modification of an initially Plankian spectrum 
(dashed line) by multiple cyclotron scattering  
for different values of $\tau_0$ and $\beta_{th}$=0.3 (from Lyutikov \& 
Gavriil 
2006).}
\label{fig:models}     
\end{figure}

%%%%%%%%%%%%%%%%%%%%%%%%%%%%%%%%%%%%%%%%%%%%%%%%%%%%%%%%%%%%%%%

%%%%%%%%%%  Figure 2 %%%%%%%%%%%%%%%%%%%%%%%%%%%%%%%%%%%%%
% For two-column wide figures use
\begin{figure}[t]
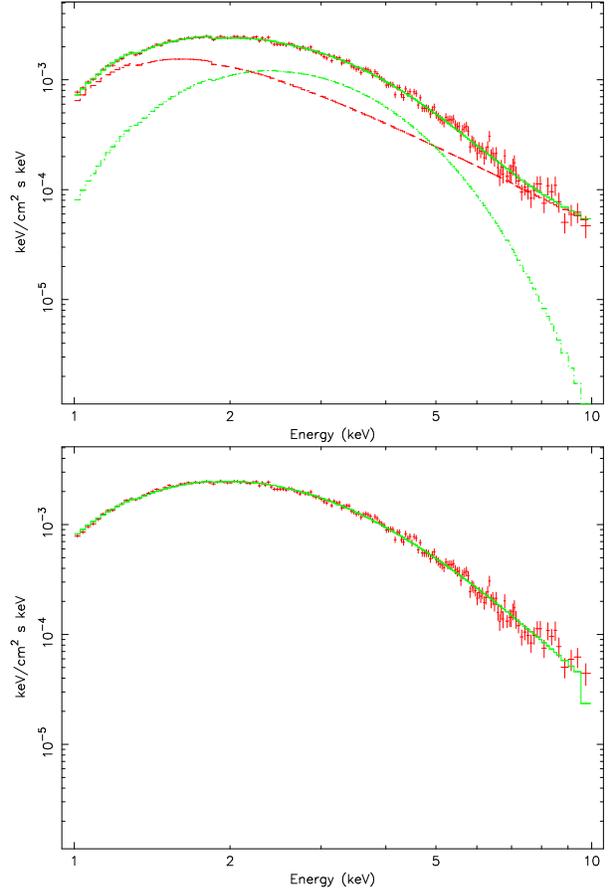

\centering
\vbox{  
\includegraphics[height=0.3\textwidth,angle=270,width=0.45\textwidth]{1048_bbpl_eufs_c.ps}
\includegraphics[height=0.3\textwidth,angle=270,width=0.45\textwidth]{1048_rcs_eufs_c.ps}}
\caption{The XMM--Newton spectrum of 1E 1048--5937 fitted with an absorbed 
power-law plus a blackbody (top panel) and with the RCS model (second 
panel).}
\label{fig:1e1048}       % Give a unique label
\end{figure}
%%%%%%%%%%%%%%%%%%%%%%%%%%%%%%%%%%%%%%%%%%%%%%%%%%%%%%%%%%%%%5

\section{The Resonant Cyclotron Scattering (RCS) model applied to 
magnetars spectra}
\label{rcsmodel}

Following the original suggestion by Thompson, Lyutikov \& Kulkarni (2002), it 
has been proposed by Lyutikov \& Gavriil (2006) that 
the presence of a warm non-relativistic plasma in the magnetosphere of
magnetars, might be responsible, through efficient resonant 
electron cyclotron scattering, of distorting the thermal X-ray emission from the 
star surface. 
In particular, if a large volume of the  neutron 
star magnetosphere of the  is partially filled by $e^\pm$-currents, the thermal (or quasi-thermal) 
cooling radiation emerging from the star surface will 
experience repeated scatterings at the 
%electron 
cyclotron resonance. 

The efficiency of the process is quantified by the 
scattering optical depth, $\tau_{res}$. Following again  Lyutikov \& 
Gavriil (2006), and 
assuming a dipolar configuration for the magnetic field, this can be 
written as 
$$ \tau_{res}= \int \sigma_{res} n dl =
 \tau_0 {( 1+ \cos ^2 \alpha)} $$ where $\sigma_{res}$ is the cross 
section for electron scattering in the magnetized regime, 
$n$ is the electrons number density, $\alpha$ is the angle between 
the photon propagation direction and the
local magnetic field, and 
$$\tau_0 = { \pi^2 e^2 n r \over
 3 m_e c \omega_B} \, .$$ 
Here $r$ is the radial coordinate,  
$\omega_B = eB/m_{e}c$ is the electron cyclotron frequency, and $B$ is the 
local value of the magnetic field. 
At variance with the case of unmagnetized Thomson scattering, in presence 
of a magnetic field $\sigma_{res}$ is resonant and is given by 
$$\sigma_{res} = {\sigma_T \over 4} { (1+ \cos ^2 \alpha) \omega^2
\over ( \omega - \omega_B)^2 + \Gamma^2/4} \, ,$$
where $\Gamma = 4 e^2 \omega_B^2 / 3 m_e c^3$ is the natural width of
the
 first cyclotron harmonic and $\sigma_T$ is the cross-section for

 unmagnetized Thomson scattering. The crucial point is that, in
this model, the plasma distribution is spatially extended. Since the
value of the resonant energy depends on 
 the local value of the
magnetic field, repeated scatterings of photons could lead to the
formation of a high energy tail instead of a  narrow line. At
relatively large distances from the surface, the magnetic  field
drops to typical values such that scattering is resonant for  soft
X-ray photons, below $\sim 10$~keV. On the other  hand, at these
energies the resonant scattering optical
 depth greatly exceeds that
for
 Thomson scattering, $\tau_{T} \sim r n
\sigma_T$,

$$ { \tau_{res} \over \tau_{T} } \sim { \pi \over 8} 
{ c \over r_e \omega_B} \sim 10^5  
\left( \frac{1\ \mathrm{keV}}{\hbar \omega_B} \right) $$
where $r_e = e^2 /mc^2$ is the classical electron radius. This
implies that even a relatively small amount of
 plasma suspended in
the magnetosphere of the neutron star may
 considerably modify the
emergent spectra. There is now increasing support 
 to the idea that the
magnetar magnetospheres 
 might be twisted (Thompson, Lyutikov \&
Kulkarni 2002), in which case
 they may be partially filled by 
 $e^-$
currents with densities well in excess of the Goldreich-Julian
density
 (which is expected in the case of a simple dipolar
configuration, see 
 Goldreich \& Julian 1968). 

 A further effect
is related to the fact that the spatial region, over
 which such
currents are distributed, is permeated by a highly inhomogeneous 
magnetic field. Although for soft X-ray photons the scattering conserves energy 
in the electron 
 rest frame, thermal/bulk electron
motion produces an energy transfer 
 in the observer rest
frame. Multiple resonant 
 cyclotron scatterings of thermal
 photons
will, on average, up-scatter the transmitted radiation, resulting
 in
the formation of a non-thermal spectrum. 
 The emerging spectrum will
then appear distorted and, as it is the case
 for thermal non-magnetic
comptonization, its shape may be well 
 represented by the
superposition of thermal and non-thermal components (see
Fig.\ref{fig:models}). 

Here we present an application of these
models to magnetar spectra. 
 Theoretical models are computed as in
Lyutikov \& Gavriil (2006), under 
 the assumption that scattering
occurs in a non-relativistic, static, warm medium and neglecting
electron recoil, i.e. in the limit 
 $\beta_{th} < 1$ (where
$\beta_{th}$, a model parameter, is the thermal velocity of electrons
in 
 units of the speed of light) and 
 for low-energy photons,
$\epsilon \ll mc^2$. 
 Within the range of parameter values allowed by
the numerical code, the non-thermal effects are 
 most prominent for
models in which the plasma is mildly 
 relativistic, $\beta_{th}
\ltsim1$, and the optical depth is reaching its boudary value
$\tau_{res}\sim 10$.

%%%%%%%%%%%%%%%%% Table 1 %%%%%%%%%%%%%%%%%%%%%%%%%%%%

\begin{table} \caption{
Best fit values of the spectral parameters obtained by fitting the 
XMM-Newton spectrum of 1E 1048--5937 with the BB+PL and an RCS model
(for more details about this observation see Mereghetti et
al.~2004). Errors are at 1$\sigma$ confidence level, and the reported
flux is absorbed and in the 2--10\,keV energy range.}

\centering
\label{tab:1}

\begin{tabular}{lcc}
\hline\noalign{\smallskip}

& BB + PL & RCS model \\
\hline\noalign{\smallskip}
  N$_{H}$ ($10^{22}$cm$^{-2}$) & 1.12$^{+0.04}_{-0.02}$ & 0.49$^{+0.01}_{-0.02}$ \\
kT (keV) & 0.64$^{+0.01}_{-0.01}$ & 0.53$^{+0.01}_{-0.01}$ \\

BB~norm &  1.03$^{+0.02}_{-0.01}\times10^{-4}$ & -- \\
$\Gamma$ & 3.3$^{+0.2}_{-0.1}$ & --  \\
PL~norm & 1.10$^{+0.13}_{-0.04}\times10^{-2}$ & -- \\

$\beta_{th}$ &  -- & 0.41$^{+0.01}_{-0.02}$ \\
$\tau_0$ & -- &  1.9$^{+0.1}_{-0.1}$ \\

RCS~norm & --   & 0.22$^{+0.01}_{-0.03}$ \\
$\chi^2_{\nu}$ &  0.99 & 1.08 \\
Flux (erg\,s$^{-1}$\,cm$^{-2}$) &7.9$^{+0.1}_{-0.1}\times10^{-12}$ & 7.9$^{+1.5}_{-1.7}\times10^{-12}$ \\
\tableheadseprule\noalign{\smallskip}
\noalign{\smallskip}\hline
\end{tabular}
\end{table}

%7.8625E-12 -  8.0818E-
%%%%%%%%%%%%%%%%%%%%%%%%%%%%%%%%%%%%%%%%%%%%%%%%%%%%%%%%%%%%%%%%%%%

\section{The RCS XSPEC model}
\label{sec:xspec}

In order to perform a quantitative comparison between fits to magnetar 
spectra performed with the model described above and with the 
canonical blackbody plus power-law model, we developed a code to 
upload the RCS model into XSPEC.  We created a grid of intensity tables 
(through a Monte-Carlo simulation) for a set of values of the three model 
parameters, i.e. $\beta_{th}$, $\tau_0$ and $T$, where $T$ is 
the temperature of the 
seed thermal surface emission (assumed to be a blackbody). The parameters 
ranges used were $0.1<\beta_{th}<0.5$ (step 0.1), $1 < \tau < 10$ (step 1) 
and $0.1$~keV$<T<3$~keV (step 0.05\,keV). For each model, the spectrum has 
been computed in the energy range 0.1--12\,keV (step 0.05\,keV). 
The final XSPEC {\it atable} spectral model has therefore four parameters, the 
three listed above plus the last one being the normalization constant, which are 
simultaneously varied during the spectral fitting following the standard
$\chi^2$ minimization technique.

Note that a previous attempt to apply the Lyutikov \&
Gavriil (2006) model to spectral data was presented by the authors 
themselves. 
However, in that case  $\beta_{th}$ and $\tau_0$ were fixed a priori 
during each fit and eventually varied manually. The fit itself was done 
only by varying $T$ and the normalization factor, while in this case we 
are able to fit simultaneously all the four parameters, resulting in 
a more precise determination of their best fitting values and 
in a more reliable $\chi^2$ determination.

%%%%%%%%%%  Figure 2 %%%%%%%%%%%%%%%%%%%%%%%%%%%%%%%%%%%%%
% For two-column wide figures use
\begin{figure}
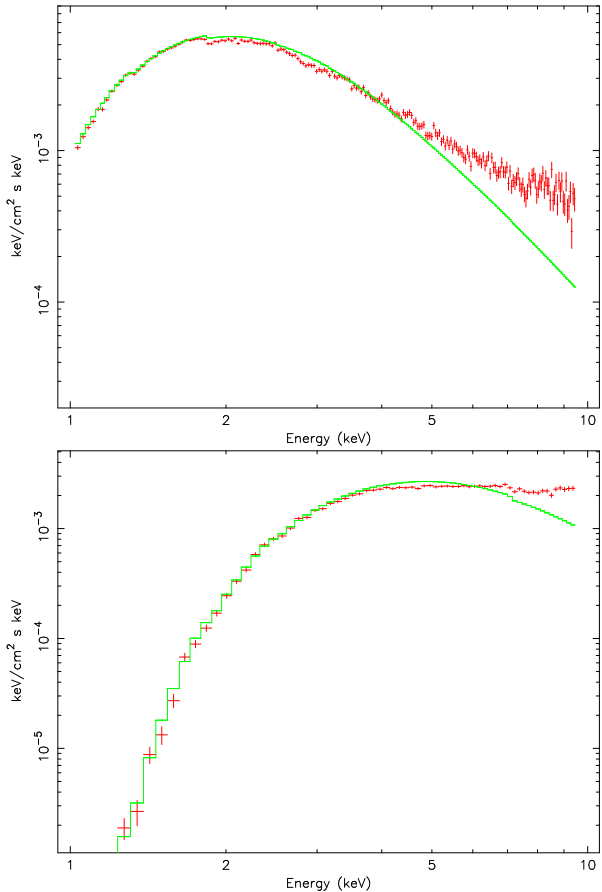

\centering
\vbox{  
\includegraphics[height=0.3\textwidth,angle=270,width=0.45\textwidth]{1708_rcs_eufs_c.ps}
\includegraphics[height=0.3\textwidth,angle=270,width=0.45\textwidth]{1806_rcs_eufs_c.ps}}
\caption{XMM--Newton spectra of 
1RXS\,J1708--4009 (top panel) and  SGR\,1806--20 (bottom panel) 
fitted with he RCS
 model. The black-body plus
power-law fits might be found in Rea et al. (2005a) and Mereghetti et al.
(2005).}
\label{fig:rcs_try}       % Give a unique label
\end{figure}
%%%%%%%%%%%%%%%%%%%%%%%%%%%%%%%%%%%%%%%%%%%%%%%%%%%%%%%%%%%%%

\section{Preliminary Results}
\label{sec:discussion}

Here we present the preliminary results of the spectral modelling of 
magnetars with the Resonant Cyclotron Scattering model
(RCS). We concentrate on the soft X-ray spectra (1--10~keV) 
of three sources, 1E 1048--56, 1RXS J1708--4009 and SGR 1806--20, and we 
use data obtained with the XMM-Newton satellite. 
Detailed information about the observations, data reduction and analysis 
can be found in Mereghetti et al.~(2004), Rea et
al.~(2005a) and Rea et al.~(2005b), for the three sources, respectively.

The fit of the X-ray spectrum of 1E\,1048--56 is presented in 
Figure\,\ref{fig:1e1048} and Table\,\,1. In this case we find that the RCS model 
is successful in reproducing the data,
as well as the canonical blackbody plus power-law model (BB+PL). The 
two fits have the same number of degrees of freedom. The value of the column 
density  found with the RCS model is lower (although both consistent with Durant \& vanKerkwijk~2006), 
while the temperature of the thermal component $T$ and the total flux are  
consistent with that found with the BB+PL modelling. These results are in 
agreement with the previous findings by Lyutikov \& Gavriil (2006), who 
presented a similar attempt by using Chandra data and varying manually 
some of the model parameters to reproduce the spectrum. 

As mentioned before, the technique used here is more reliable in
providing a determination of the best fitting parameters. In this
case, the source spectrum can be reproduced by invoking
scattering by $e^-$ with thermal velocity $\beta_{th}=0.41$ and
optical depth $\tau_0=1.9$.

On the other hand, the cases of 1RXS\,J1708--4009 and SGR\,1806--20 
are more problematic. These sources have much harder 
X-ray spectra (in the 1--10~keV band), that cannot be accounted for by the   
RCS model alone (see Fig.\,\ref{fig:rcs_try}) at least when the 
spectral deformation  resulting from the resonant scattering 
process is computed in the simple way proposed by Lyutikov \& Gavriil 
(2006). 

However, it is  worth noting that, at variance with 
1E\,1048--56, these two sources exhibit strong high-energy power-law 
tails which extend up to the $\gamma$-rays (Kuiper et
al. 2006; Molkov et al. 2005; Mereghetti et al. 2005b). Motivated by 
that, we fitted the whole 1--50\,keV spectra  with the RCS model plus a
power-law. For both sources, we obtained a good fit with a 
reduced $\chi^2$ of 1.08 and 1.1, respectively. Although 
extremely promising, these analysis 
are very preliminary and they still require a careful assessment of 
intercalibration issues which arise when different detectors are used.
Therefore, we do 
not discuss the results in more detail nor we present here best fitting 
parameter values. These results will 
be presented in more details in a forthcoming paper (Rea et al.~2006, in prep.).

\section{Conclusion}
The preliminary results reported here show that the X-ray
spectra of the AXP 1E\,1048--56 can be successfully reproduced 
 by the
RCS model.  The quality of the fit is comparable to that 
 obtained by
using the canonical blackbody plus power-law model, with the
advantage that the RCS model spectra provide a better
physical understanding of the presence of the thermal/ non-thermal
components in the observed spectrum. On the other hand, this is not
totally unexpected  since this source has a non-thermal tail which is
relatively soft with  respect to other AXPs, and a similar result has
already been found by  Lyutikov \& Gavriil, although their fitting
technique was less accurate. For the other two sources
considered here, 1RXS\,J1708--4009
 and SGR\,1806--20, we find that the
RCS model alone cannot account for the relatively hard non-thermal
component observed below $\sim$10~keV.

% {\bf This is not entirely unexpected, since the semi-analytical 
%model of Lyutikov \& Gavriil  (2006) {\it assumes} scattering plasma
%is non-relativistic and thus is expected to be unaplicable to very
%hard spectra}.

On the other hand, in both cases a composite model consisting of a RCS
model 
 plus a power-law can fit the X-ray spectra in the 1--100~keV
band. This may  suggest that, in the 6--10\,keV energy range, the
X-ray  emission of these sources is contaminated by the
 hard 
non-thermal component which has beed recently observed with INTEGRAL 
 at
much higher energies (Kuiper et al. 2006). A similar hard tail has not
been observed, at least so far, in the case of 1E\,1048--56. A detailed
paper with a systematic application of the RCS model to the whole
class of magnetars and a study of their spectral and flux variability
in this scenario is in preparation, accompanied by a  more complete
discussion (Rea et al.~2006, in prep.).
\begin{acknowledgements}
We thank Valentina Bianchin and Gavin Ramsay for their kind help in
building the XSPEC RCS model. We also thank Fotis Gavriil for having
kindly allowed us to look into his preliminary model, Andrea Tiengo,
Gianluca Israel and the anonymous referee for useful comments. NR
thanks the Mullard Space Science Laboratory, where this work was partially done,  for the warm hospitality.

\end{acknowledgements}


\begin{thebibliography}{}


\bibitem{} 
Camilo, F., Ransom, S., Halpern, J., et al. 2006, Nature submitted, preprint (astro-ph/0605429)
\bibitem{} 
Chatterjee, P., Hernquist, L. \& Narayan, R. 2000, ApJ, 534, 373

\bibitem{} 

Durant, M. \& vanKerkwijk, M.H., 2006, ApJ in press, astrp/0606604

\bibitem{}
Duncan, R.C. \& Thompson, C.  ApJ {\bf 392}, L9 (1992)

\bibitem{} 
Goldreich, P. \& Julian, W.H. ApJ {\bf 157}, 869 (1969)
\bibitem{}
Hulleman, F., van Kerkwijk, M. H. \& Kulkarni, S. R. Nature {\bf 408}, 689 (2000)
\bibitem{} 
Kuiper, L., Hermsen, W. \& Mendez, M.  ApJ {\bf 613}, 1173 (2004)	
\bibitem{} 
Kuiper, L., Hermsen, W., den Hartog, P. R., Collmar, W. ApJ {\bf 645}, 556 (2006)
\bibitem{}
Lyutikov, M. \& Gavriil, F.  MNRAS {\bf 368}, 690 (2006)

\bibitem{}
Mereghetti, S., Tiengo, A., Stella, L., et al. ApJ {\bf 608}, 427 (2004)	
\bibitem{}
Mereghetti, S., Tiengo, A., Esposito, P., et al. ApJ {\bf 628}, 938 (2005a)
\bibitem{}
Mereghetti, S., Gotz, D., Mirabel, I. F., Hurley, K.  A\&A {\bf 433}, L9 (2005b)
\bibitem{}
Molkov, S., Hurley, K., Sunyaev, R., et al.  A\&A {\bf 433}, L13 (2005)
\bibitem{}
Rea, N., Oosterbroek, T., Zane, S., et al. MNRAS {\bf 361}, 710  (2005a)

\bibitem{}
Rea, N., Israel, G.L., Covino, S., et al.  The Astronomer's Telegram, \#645 (2005b)

\bibitem{}
Tiengo, A., Mereghetti, S., Turolla, R., et al.  A\&A {\bf 437}, 997 (2005)
\bibitem{} 
Thompson, C. \& Duncan, R.C.  ApJ {\bf 408}, 194 (1993)

\bibitem{} 
Thompson, C. \& Duncan, R.C. ApJ {\bf 473}, 322 (1996)

\bibitem{} 
van Paradijs, J., Taam, R.E. \& van den Heuvel, E.  A\&A {\bf 299}, L41 (1995)
\bibitem{}
Woods, P.~M. \& Thompson, C.,  preprint (astro-ph/0406133) (2004)




%\bibitem{Ref1}
%Author, I.: Article title. Journal Title-Abbreviated {\bf Vol}, pp--pp (year)
% Format for books
%\bibitem{Ref2}
%Author, I., Smith, J.: Book Title. Publisher, Place (year)
% Format for proceedings
%\bibitem{Ref3}
%Author, I., Smith, J.: Paper title. In: Editor, A. (ed.) Proceedings
%Title, Location, Date, pages. Publisher, Place (year)
% etc

\end{thebibliography}
\end{document}